\documentclass[aps,amsfonts,pra,twocolumn]{revtex4}

\usepackage{subfigure}
\usepackage{textcomp}
\usepackage{graphicx}
\usepackage{amssymb}
\usepackage{amsmath}
\usepackage{bm}
\usepackage{amsmath, amsthm, amssymb}
\usepackage{dsfont}
\usepackage{url}

\def\beq{\begin{equation}}
\def\eeq{\end{equation}}
\def\bea{\begin{eqnarray}}
\def\eea{\end{eqnarray}}

\usepackage[usenames, dvipsnames]{color}

\begin{document}

\title{Instability of many-body localized systems as a phase transition in a nonstandard thermodynamic limit}

\author{Sarang Gopalakrishnan}
\affiliation{Department of Physics and Astronomy, CUNY College of Staten Island, Staten Island, NY 10314 USA}
\affiliation{Physics Program and Initiative for Theoretical Sciences, CUNY Graduate Center, New York, NY 10016 USA}
\author{David A. Huse}
\affiliation{Department of Physics, Princeton University, Princeton, NJ 08544 USA}

\begin{abstract}
The many-body localization (MBL) phase transition is not a conventional thermodynamic phase transition.  Thus to define the phase transition one should allow the possibility of taking the limit of an infinite system in a way that is not the conventional thermodynamic limit.  We explore this for the so-called ``avalanche'' instability due to rare thermalizing regions in the MBL phase for quenched-random systems in more than one spatial dimension, finding an unconventional way of scaling the systems so that they do have a type of phase transition.  These arguments suggest that the MBL phase transition in systems with short-range interactions in more than one dimension is a transition where entanglement in the eigenstates begins to spread in to some typical regions: the transition is set by when the avalanches {\it start}.  Once this entanglement gets started, the system does thermalize.  From this point of view, the much-studied case of one-dimensional MBL with short-range interactions is a special case with a different, and in some ways more conventional, type of phase transition.
\end{abstract}

\maketitle

\section{Introduction}

Isolated quantum systems can exhibit the phenomenon of many-body localization; many-body localized (MBL) systems do not thermally equilibrate themselves, and can preserve \emph{local} memories for arbitrarily long times~\cite{basko_metalinsulator_2006, gornyi_interacting_2005, oganesyan_localization_2007, znidaric_many-body_2008, nandkishore_mbl_2015, serbyn_universal_2013, serbyn_local_2013, hno, Schreiber15}. A stable MBL ``phase'' has been proved to exist, under certain minimal assumptions, for one-dimensional spin chains evolving under strongly disordered local Hamiltonians~\cite{jzi, jzi2}.  In this phase, all eigenstates are localized in the sense of having area-law entanglement~\cite{oganesyan_localization_2007, bauer_area_2013}, and there is a complete set of spatially quasi-local, mutually commuting integrals of motion (called LIOMs or l-bits)~\cite{serbyn_local_2013, hno}. Whether MBL exists in more general systems---for instance, in higher dimensional systems~\cite{Hild16, Bordia16, bordia2d17, wahl2017signatures}, systems with long-range interactions~\cite{burin1998, burin2006, ylg, burin2015, burin2015b, gornyi2017spectral, PhysRevB.93.245427, tikhonov2018}, or systems subject to quasiperiodic rather than random potentials~\cite{iyer_many-body_2013, Schreiber15, lueschencrit16, ksh, sdp, vznidarivc2018interaction}---is less firmly established. 
Indeed, a recent argument~\cite{drh, ldrh} suggests that when one takes the thermodynamic limit in the conventional fashion, the MBL phase in disordered systems is generically unstable to rare locally thermalizing regions in dimensions $d > 1$, as well as in one dimension for interactions that decay slower than exponentially with distance.  This is the so-called ``avalanche'' instability where a slow avalanche of entanglement spreads from these rare regions through typical regions, and thermalizes the entire system.  
Nevertheless, there is experimental evidence for a sharp crossover to very slow dynamics in certain two-dimensional systems~\cite{Hild16, bordia2d17}, even if true MBL is in some sense absent. 

In the present work, we assume that this ``avalanche'' instability exists as proposed in Refs.~\cite{drh, ldrh}, and discuss its implications for the MBL phase transition as a function of sample size and/or time.  We consider spin models on $d$-dimensional lattices, with random fields of strength of order one and spin-spin interactions of strength $J$.  {If these are considered as particle models, the random potential is of order one and both the hoppings and the interactions are of strength $J$.}  We consider short-range interactions and interactions that fall off with distance $r$ as slowly as a power law: $J(r) \sim J/r^{\alpha}$.  For finite-size systems of linear size $L$ we estimate the interaction $J_c(L)$ where the eigenstates change from localized to thermalized.  For finite $L$ this change is, of course, only a crossover, but we argue that under the appropriate scaling this crossover sharpens up to a type of phase transition in the limit $L\rightarrow\infty$.  Similarly, we can instead consider an infinite system and look at the behavior as a function of time $t$, asking if the time dependence of the operator spreading in the system is that characteristic of the MBL phase or characteristic of the thermal phase, and define a $J_c(t)$ that way.

This unconventional way of taking the $L \rightarrow \infty$ limit can be motivated as follows:  Recall that in systems with sufficiently long-range interactions, even conventional thermodynamic phase transitions take place at an interaction strength that decreases to zero as the system size is taken to infinity. In these cases it is conventional to scale the interaction with system size so that its contribution to, say, the total free energy remains extensive as $L \rightarrow \infty$. For the study of the MBL transition what is important is not thermodynamics, but instead the dynamics of entanglement. The dynamics are more sensitive to interactions than is the thermodynamics, so we find that in order to have a MBL transition, in most cases we need to scale the interactions to zero more strongly than we would if we were only trying to capture the thermodynamic effects of the interactions.

Looking across the range of cases that we consider, varying $d$, $\alpha$, $L$ and $t$, we find that there are three different types of estimates of the transition $J_c$, each applicable in a different regime:  One estimate may be termed ``Fock-space localization'', and ignores rare region effects \cite{basko_metalinsulator_2006}.  A second estimate, that we introduce here, is where the avalanches get {\it started}.  And the third estimate is that used by Refs.~\cite{drh, ldrh}, which is to assume that the avalanches do get started and ask when the avalanches spread without limit through the entire system.

We find that the MBL transition in the large $L$ or large $t$ limit is set by when the avalanches get started for systems with short-range interactions in $d>1$, as well as more generally for systems with longer range interactions, including power-law interactions with $\alpha>2d$ that do not decay too slowly with distance.  These are all systems where a sufficiently large avalanche will not stop, so once the avalanche gets started, the system does thermalize.  In these cases we estimate the probability of there being a sufficiently large locally thermalizing rare region to start an avalanche.  This gives estimates of the transition  behaving as $\log(J_c(L)) \sim -(\log L)^{1/3}$ and $\log(J_c(t)) \sim -(\log (\log t))^{1/3}$ for systems with short-range interactions.  Thus to obtain a phase transition, we need to scale the interaction down to zero as we increase $L$ or $t$ to infinity.  And we find that as we take this unconventional scaling limit the transition does become sharp, so this is not ``just a crossover''.

When power-law long-range interactions become even longer, so $\alpha<2d$, then delocalization happens due to long-range system-wide resonances between typical spins before rare region effects become important.  So in this regime the Fock-space localization estimate of the transition that ignores rare region effects is correct~\cite{burin2015b, tikhonov2018, plhc}.  This gives estimates $J_c(L) \sim L^{-(2d-\alpha)}$ and $J_c(t)\sim t^{-(2d-\alpha)/(2d)}$ (up to multiplicative logarithmic corrections) that depend more strongly on $L$ and $t$.  In both of these regimes that we have discussed so far, to obtain a sharp MBL phase transition we must scale the interactions to zero more rapidly with increasing $L$ than is done in the conventional thermodynamic limit.  In this sense, at the transition the interactions are thermodynamically irrelevant but dynamically relevant, as emphasized for a somewhat different class of MBL transition in Ref.~\cite{plhc}.

The case of $d=1$ with short-range interactions is special, since here the transition in the limit of large $L$ and $t$ is not set by when the avalanches get started, but instead by when they ``run away'' without being stopped by the typical disordered regions.  In this case this actual MBL phase transition occurs at a nonzero $J_c$ in the conventional thermodynamic limit.  But even here we can consider the other estimates of the transition for smaller finite $L$ or $t$ on the scale of numerical studies or experiments.  We suggest that it might be that for many of the models that have been studied, the {apparent $J_c(L)$ that is seen may be mostly set by either Fock-space localization in typical regions or the threshold for {\it starting} avalanches. 
Then as $L$ is increased, a given sample becomes more likely to have a large thermal inclusion, so the avalanche-start estimate $J_c(L)$ decreases with increasing $L$ and eventually drops well below} the actual phase transition that is set by when the avalanches run away.  But this crossover to the asymptotic critical behavior might happen at very large $L$ for some, and possibly all, of the models that have been studied so far.

This paper is organized as follows. In Sec.~\ref{models} we specify the models to be considered and summarize our main results. In Sec.~\ref{higherd} we discuss short-range systems in dimensions greater than one. In Sec.~\ref{fastpower} we discuss the case of sufficiently rapidly decaying power laws. In Sec.~\ref{1dweak} we discuss one-dimensional systems with short-range interactions, {including particle models where we allow the hoppings to be strong compared to the interactions.}  Finally, in Sec.~\ref{disc} we discuss some implications of these results. 

\section{Background, models, and main results}\label{models}

In this work we are concerned with general Hamiltonians of the form
\beq
H = \sum_i h_i O_i + J \sum_{ij} \phi_{ij} P_{ij},
\eeq
where the $h_i$ are random fields that are of order one, the $O_i$ are single-site operators that are of order one, 
the $\phi_{ij}$ are (potentially random) coefficients that are of order one at distance one and fall off in a prescribed way with the distance $r_{ij}$ between the two sites $i$ and $j$, and the $P_{ij}$ are two-site operators with norm of order one that do not commute with $O_i$ or $O_j$.  We ``tune'' through the MBL transition by varying $J$. 

In general the local Hilbert space could have more than two states per sites, but for simplicity and specificity we will consider the case of two states per site.  The generalization to $q$-state systems with $q > 2$, or operators that act on more than two sites, is fairly straightforward, and will only affect some prefactors in the estimates below.  In general we could also consider the behavior as a function of temperature, but again for simplicity and specificity we restrict our attention to infinite temperature, where the full state space is ``in play''.  
For concreteness, we refer to the local degrees of freedom as ``spins''.  We will consider power-law, stretched exponential, exponential, and strictly short-ranged forms for $\phi_{ij}$.  In the power-law cases, we take the interactions to fall off as $1/r^{\alpha}$; in the following discussion, $\alpha$ will always denote this exponent.  This Hamiltonian may be time-independent, or it may be periodic in time with a period of order one, thus making a Floquet system~\cite{Ponte15, abanin_theory_2014}.  In the latter case, when we refer to ``energy'', it means the pseudo-energy.  Since the focus is on low-energy dynamics this is not an important distinction for the physics here.


\subsection{Background: long-range models}\label{bg}

In models with sufficiently slowly decaying power law interactions, delocalization occurs primarily through few-spin resonances and we do not need to invoke rare-region physics. The general mechanism is as follows~\cite{burin1998, burin2006, ylg, burin2015, burin2015b, gornyi2017spectral, PhysRevB.93.245427, tikhonov2018}:  Suppose the interactions fall off as $J/r^\alpha$. Then a typical spin is involved in $\sim J R^{d-\alpha}$ resonances at scale $R$. Each resonant pair of spins forms a two-level system, consisting of hybridized states.  Crucially, all such two-level systems at scale $R$ are coupled to one another by a matrix element comparable to their detuning, so an $O(1)$ fraction will inter-resonate. The number of two-level systems within a distance $R$ of a particular two-level system can be estimated as $\sim J R^{2d-\alpha}$; when $\alpha < 2d$ sufficiently large two-level systems always inter-resonate, and the system delocalizes on long enough scales. To prevent this delocalization, one must scale $J$ down with system size $L$ such that $J L^{2d - \alpha}$ always remains small, so $J_c(L) \sim L^{\alpha - 2d}$. In particular, in the all-to-all case $\alpha = 0$, the localization transition is at $J_c(L) \sim L^{-2d}$. These arguments hold up to multiplicative logarithmic corrections, which were noted in Ref.~\cite{tikhonov2018}. 

The proper scaling of the interactions in the conventional thermodynamic limit is $J(L) \sim L^{\alpha - d}$ for $\alpha < d$, while $J$ does not need to be scaled to zero for $\alpha > d$ to get an interacting thermodynamic limit.  Thus we find that the scaling that is needed to see the localization transition is such that the interactions scale to zero with increasing $L$ faster than they do in the conventional thermodynamic limit, so that their contribution to the equilibrium thermodynamics vanishes in the large $L$ limit:  The interactions are thermodynamically irrelevant, but dynamically relevant when scaled as needed to see the MBL phase transition.

\subsection{Avalanche argument}

Localization in systems with power laws $\alpha > 2d$, or short-range systems, is stable against few-spin resonances at weak interactions. However, these systems appear to be susceptible to a non-perturbative rare-region instability seeded by thermal inclusions.  Following standard usage we call this instability an ``avalanche''~\cite{drh}, although this instability is an extremely slow process, so this name is perhaps misleading.
To explain the avalanche argument we first recall that an MBL Hamiltonian, with an entirely localized spectrum, can be written in terms of its spatially quasi-local integrals of motion (l-bits), $\tau_i^z$, as follows~\cite{hno, serbyn_local_2013, jzi, pcor, varma2019}:
\beq\label{lbit}
H = \sum_i h_i \tau^z_i + \sum_{ij} J_{ij} \tau^z_i \tau^z_j + \sum_{ijk} K_{ijk} \tau^z_i \tau^z_j \tau^z_k + \ldots
\eeq
where these interactions fall off exponentially with the distance between the sites involved (specializing for now to the case where the microscopic interactions are short-range).  Further, a generic local physical operator $\tilde O_i$ can be expanded in the basis of $\tau$ operators, as $\tilde O_i = \sum_j c^{(1,i)}_{j,\alpha} \tau^\alpha_j + c^{(2,i)}_{jk,\alpha\beta} \tau^\alpha_j \tau^\beta_k + \ldots$. The coefficients $c$ also fall off exponentially with distance.  Among the coefficients at a fixed distance, those that involve flipping many l-bits may be further suppressed:
in general, the amplitude of a particular term is $\sim J^n$ where $n$ is the lowest order in which that term is generated in the locator expansion~\cite{ros2015integrals}.  

The avalanche argument proceeds as follows~\cite{drh, ldrh}:  In the MBL phase, assuming a model of uncorrelated disorder, there will be rare large regions where the disorder is locally weak (e.g., all the local fields are approximately the same).  Any such region of finite size can occur with some nonzero probability.  To address these regions, we first imagine cutting all interactions between them and the typical regions, and performing a local unitary transformation that diagonalizes the latter in the l-bit basis.  The rare thermal spots, meanwhile, are assumed to be ``good'' thermal regions, which locally obey the eigenstate thermalization hypothesis (ETH).  We then reinstate the interactions between the thermal spots and the typical regions; this corresponds to adding interactions of the form $O_{\mathrm{edge}} O_{\mathrm{th}}$, where the operator $O_{\mathrm{edge}}$ is a generic local operator (with an expansion in terms of l-bits as described above) and $O_{\mathrm{th}}$ is an operator on the thermal spot.  Spins that are a distance $x$ from the thermal spot are coupled to it via spin-flip matrix elements of $O_{\mathrm{edge}}$ that fall off exponentially with $x$.  The spin gets entangled with the thermal spot if this spin-flip matrix element coupling it to the spot exceeds the many-body level spacing of the spot.

The key observation is that spins near the edge of a sufficiently large rare spot invariably entangle with it.  Once this happens, the effective level spacing of the spot is reduced (because it is now enlarged by the spins it has ``absorbed''); farther-out spins then entangle with the spot if their matrix elements exceed this \emph{reduced} level spacing.  Asymptotically, in dimensions greater than one, a sufficiently large spot invariably induces an instability: when a spot of 
large surface area $A$ absorbs the spins that immediately surround it, its level spacing goes down by a factor of $2^{-A}$, whereas the matrix element for the next step only decreases by a factor of $\exp(-1/\zeta)$, where $\zeta$ is a decay length for the spin-flip couplings.  Thus once this ``avalanche'' starts, it will not stop and the eigenstates of the full sample will be thermal.

\subsection{Main results}

We now summarize the main results of our analysis, in the various cases considered, going from the longest-ranged to the shortest-ranged models.

\emph{Very long-range power laws $\alpha < d$}.---In this case, delocalization is due to the proliferation of few-spin resonances on the scale of the system size; these resonances are the first to proliferate. The relevant spins in a system of size $L$ are all coupled to each other with the typical coupling $J/L^\alpha$; one can thus regard these models as analogous to the all-to-all model, but with an effective coupling that renormalizes with $L$. The all-to-all model has a transition with a critical point that scales as $J_c \sim 1/N^2 \sim 1/L^{2d}$. We can put these two results together and get
\beq\label{plscaling}
J_c(L) \sim 1/L^{2d - \alpha}.
\eeq
In these models, the inverse participation ratio in Fock space remains finite throughout the localized phase. Thus the transition here is a Fock-space delocalization transition of the type discussed in Ref.~\cite{basko_metalinsulator_2006}.

Equivalently, instead of a length-dependent critical coupling we can define a \emph{time}-dependent critical coupling using the fact that a resonance at scale $L$ gives rise to hybridization on timescale $t(L) \sim L^\alpha/J(L)$. Thus, $J_c(t) \sim 1/(J_c(t) t)^{(2d-\alpha)/\alpha}$. This then implies
\beq
J_c(t) \sim 1/t^{1 - (\alpha/2d)}.
\eeq

\emph{Intermediate-range power laws, $d < \alpha < 2d$}.---In this case, Eq.~\eqref{plscaling} continues to apply for the critical coupling; however, there are some conceptual differences. In this case, isolated short-range resonances are present at some nonzero density even in the MBL phase.  A very rough estimate of the inverse participation ratio, taking into account only nearest-neighbor resonances, is 
\beq
\text{IPR} \sim \exp(-N_{\mathrm{res.}}) \sim \exp(-L^d J(L)) \sim \exp(-L^{\alpha - d}),
\eeq
which evidently vanishes when $\alpha > d$.  However, we can relate this case to a Fock-space localization problem as follows: first, we separate out the interaction into a short-range piece that is cut off at a scale that is a small but finite fraction of $L$, and a long-range piece consisting of terms with range comparable to $L$. The short-range piece can be diagonalized in terms of quasi-l-bits (which would be exact l-bits if we neglected the long-range interactions). We then express the long-range couplings in terms of these l-bits; these long-range couplings create resonances that are system-wide in scale, and can be treated as in the all-to-all case. 

\emph{Faster power laws, $\alpha > 2d$}.---In this case, few-spin resonances do not proliferate when $J$ is sufficiently small (even if we do not scale it with system size).  In the absence of avalanches, the MBL phase would be stable for small enough $J$, with algebraically localized l-bits.  However, avalanches destabilize it (as analyzed in more detail in Sec.~\ref{fastpower}); the critical coupling (or characteristic size scale) for an avalanche is given by
\beq\label{plavalanche}
J_c(L) \sim \exp(-\text{constant} \times \alpha \sqrt{\log L}).
\eeq
The avalanches in this case are different from those in short-range models, in that they can be spatially sparse clusters of spins. Using the fact that $J_c(L)$ vanishes slower than a power law of $L$, we can deduce that to leading order, $t \sim L^\alpha$ and so
\beq
J_c(t) \sim \exp(-\text{constant} \times \sqrt{\alpha \log t}).
\eeq

\emph{Short-range models in $d > 1$}.---For short-range models, we estimate the large-$L$ scaling
\beq\label{sravalanche}
J_c(L) \sim \exp[-\text{constant} (\log L)^{1/3}].
\eeq
Once again, $J_c(L)$ vanishes rather slowly as $L \rightarrow \infty$, so the length-time relation is $L \sim \log t$. Plugging this in, we find that
\beq
J_c(t) \sim \exp[-\text{constant} (\log \log t)^{1/3}].
\eeq
Thus the critical coupling scales to zero extremely slowly as a function of time, and on experimentally realistic timescales it may look essentially time-independent.

\emph{Width of transition regions}.---We now discuss the width of these transitions:  Does the finite-$L$ crossover sharpen up and become a phase transition in the limit of large $L$, or does it remain a crossover? 

First, when $\alpha < 2d$, the transition essentially occurs~\cite{tikhonov2018} when each sample contains on the order of one resonance on the scale $L$.  In the ``dynamic'' limit we are considering, to leading order we scale $J_c \sim 1/L^{2d - \alpha}$ to keep the expected number of resonances in a sample of size $L$ fixed and $O(1)$.  In this leading-order estimate, we find that the width of the crossover region---in which the probability of having an order one number of resonances in a given sample differs appreciably from zero or one---remains a finite fraction of $J_c$ regardless of $L$.  To understand the scaling of the transition region, therefore, we must go beyond this leading-order analysis.  As this is not the main focus of the present work, we simply remark that numerical scaling studies of this case~\cite{tikhonov2018} observe that the transition sharpens as $L$ is increased.

We now turn to the case when avalanches are involved.  Here, we find that the critical region sharpens as $L$ (or $t$) is increased. For concreteness we consider the short-range case in $d > 1$. Here the probability of an avalanche in a system of size $L$ at coupling $J$ is given by
\beq
P(J, L) \sim L^d \exp(-\text{constant} \times |\log^3 J|). 
\eeq
We fix $L$ and look for couplings $J_\pm$ for which the probability of an avalanche is (e.g.) $1/2 \pm w$ respectively (one can think of these, e.g., as quartiles ($w=1/4$) of the probability distribution). We find that $\log J_+ \sim [\log\{ (1/2 + w) L^{-d} \}]^{1/3}$ and similarly for $\log J_-$. The transition width can be estimated as 
\beq
(\delta J_c)/J_c^{\mathrm{typ}} \approx \log J_+ - \log J_- \sim (\log L)^{-2/3}.
\eeq
Thus the transition point $J_c(L)$ becomes sharply defined as $L \rightarrow \infty$ in this scaling limit: the probability changes from near zero to near one on a change in $J$ that is small compared to $J_c$ itself.  We will see below that in this same sense the time-dependent $J_c(t)$ also becomes asymptotically sharply defined.  However, one can check that $L_c(J)$ and $t_c(J)$ do \emph{not} become sharply defined in the same way.

\section{Higher-dimensional short-range systems}\label{higherd}

We now consider $d$-dimensional systems with short-range interactions, for $d>1$.  For small enough systems or short enough times, avalanche physics may be unimportant and the crossover from localization to thermalization may be best estimated with a Fock-space localization estimate.  But if the sample size $L$ and the time $t$ are large enough, avalanches will happen at the Fock-space localization estimate of the transition~\cite{drh,ldrh}, and we need to instead estimate at what coupling these avalanches start.  This section examines this estimate in more detail, always assuming that $L$ and $t$ are large enough so that the transition is indeed set by the avalanches getting started.

\subsection{Constructing the critical thermal inclusion}
 
We anticipate that the shape of a thermal inclusion is relatively unimportant to its ``potency'', and construct inclusions as follows:  The first site is ``free'', meaning unrestricted.  The second site has to neighbor the first site and in order to be resonant with it, the two spins must have have their $|h_i|$'s within $J$ of each other, similarly with the third site, etc. For a given lattice, 
a rough estimate is that the density of such thermal inclusions of $\ell$ spins is $(zJ)^{(\ell -1)}$ (where $z$ need not be an integer).  A typical such inclusion will have a fractal dimension between $1$ and $d$.


The inclusion consists of spins which are all at approximately the same field $|h_i|\simeq h$; therefore, the coupling $J$ and this average field $h$ set the inclusion's internal energy scales.  In particular, its local spectral functions are concentrated within $\sim J\sqrt{\ell}$ of energies $0$ and $\pm h$.
We anticipate that $J$ will eventually be taken to zero as we take the limits of large $L$ or $t$, and for small enough $J$ the bath has a small bandwidth, so the arguments of Ref.~\cite{gopalakrishnan2014mean} apply.  Also, we will obtain $\ell \sim \log^2 J$, which allows us to drop some factors of $\ell$ in getting the leading behavior.

Consider a typical spin near the inclusion.  The most likely way for this spin to get entangled with the inclusion in an eigenstate is via a high-order rearrangement of spins; to get such a rearrangement with energy within $\alt J$ of resonance with the inclusion one needs to include of order $\sim |\log J|$ other spins in the process.  At this scale the matrix element is $\sim 2^{-\log^2 J}$, so vanishes with decreasing $J$ faster than any power of $J$.  We require this to be smaller than the level spacing of the inclusion, which is $\sim J2^{-\ell}$, for the inclusion not to absorb this multi-spin rearrangement.  
To stop this inclusion from growing and thus initiating an avalanche we need the inclusion to be off-resonant with each of the possible multi-spin rearrangements that it can couple to in the surrounding typical spins.  The number of these possible $|\log J|$-spin resonances grows with decreasing $J$ no faster than a power of $J$, so the factor due to all these possible resonances can be neglected relative to the stronger $J$-dependence of the matrix element.
Thus we need $2^{-\log^2 J} \alt 2^{-\ell}$ to prevent this inclusion from initiating an avalanche.  To leading order, the critical size for a rare thermal inclusion is thus $\ell \simeq \log^2 J$ spins. 

Since our approach to finding the most potent inclusion for starting an avalanche is not fully systematic, one might worry about whether there is some more probable inclusion that we have neglected.  In order to have a different scaling, it seems like the inclusion would need to either have a bandwidth that depends on $J$ more slowly than a power law, and/or need to be able to couple to a number of possible multispin rearrangements that grows faster than $\sim 2^{\ell}$.  With only $\ell$ spins in the inclusion, neither of these scenarios seem possible.  

\subsection{Location and width of the transition}

\begin{figure}[!t!b]
\begin{center}
\includegraphics[width = 0.45 \textwidth]{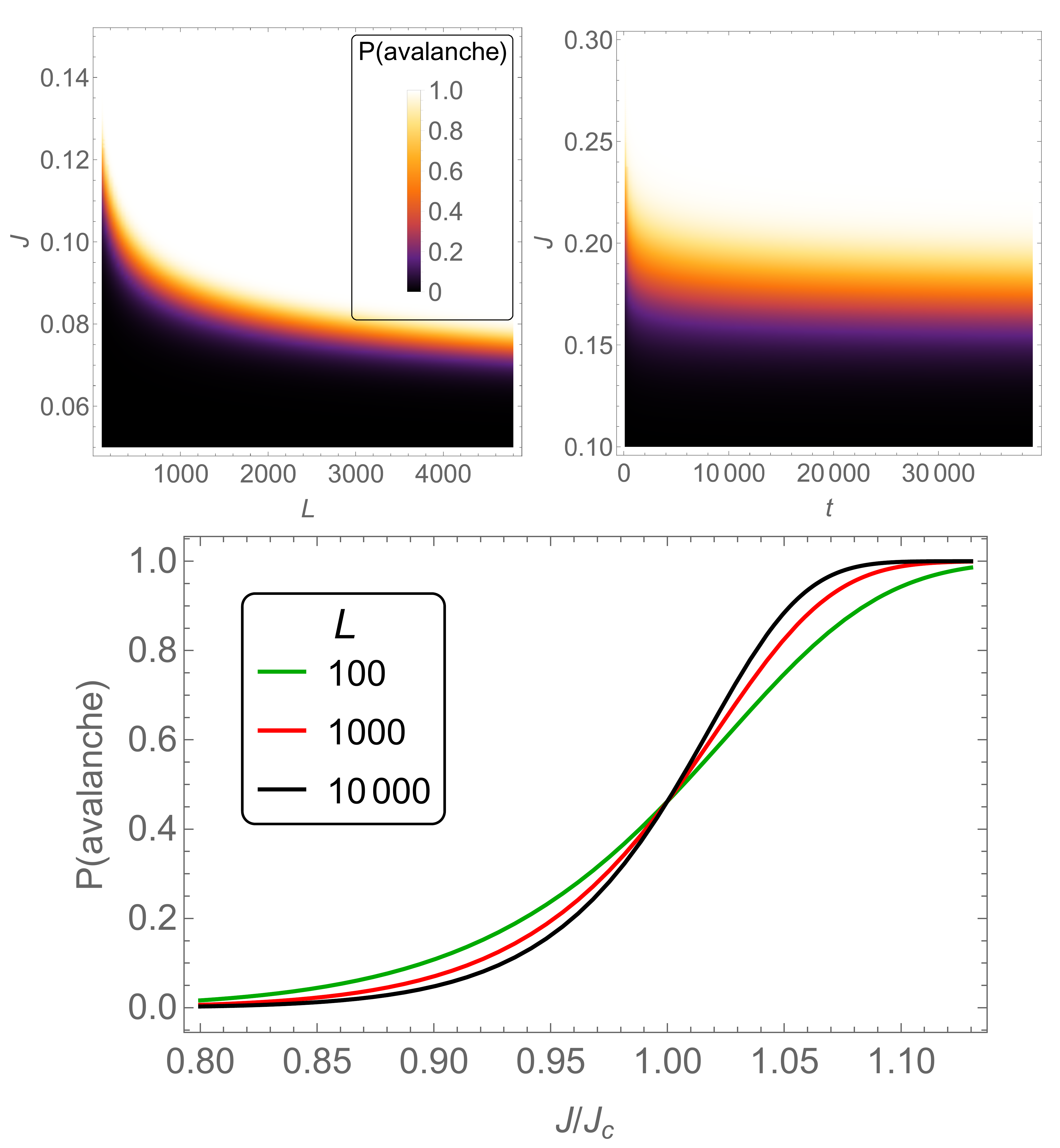}
\caption{For systems in dimensions $d>1$ with short-range interactions:  Upper left: probability of a sample of size $L$ containing an avalanche, as a function of $J$ and $L$.  Upper right: probability of a typical degree of freedom being incorporated into an avalanche at time $t$, as a function of $J$ and $t$.  Lower panel: probability of a sample of size $L$ containing an avalanche vs. rescaled coupling $J/J_c(L)$ at various $L$. Note the gradual sharpening of the transition as $L$ is increased. These plots illustrate the behavior of the functions~\eqref{scrit2}, \eqref{critt}; however, the quantitative values other than $J/J_c$ are only schematic, as we have set various undetermined constants to unity.}
\label{fig1}
\end{center}
\end{figure}

For a given linear system size $L$, the probability of a thermal inclusion that is large enough to start an avalanche being present is 
\beq\label{scrit2}
P(L, J) \sim L^d J^{-\log^2 J} ~.
\eeq 
The critical coupling then obeys 
\beq\label{srcrit}
|\log J_c(L)| \sim [ d \log L]^{1/3}.
\eeq

For any finite $L$ the change at $J_c(L)$ is not sharp, it is only a crossover, as is almost always the case for phase transitions in finite systems.  But we can ask if this crossover sharpens up to a phase transition as we take the limit of large $L$.
The probability that a finite system of $L^d$ spins has avalanched is $P(L, J) \sim L^d \exp[- |\log^3 J|]$.  The width of the transition region can be estimated by estimating the change, $\delta J$, in the coupling $J$ between, say, the values where this probability is 1/4 and 3/4.  
By this measure the transition does indeed sharpen up as
\beq
\frac{\delta J}{J_c(L)} \sim (\log L)^{-2/3}
\eeq
as the limit of large $L$ is taken.
We note that although $J_c(L)$ becomes asymptotically well-defined, repeating the analogous calculation shows that the $L$-driven crossover at fixed $J$ remains broad: there is no sharply defined critical size $L_c(J)$ at which avalanches appear. 

\subsection{Finite-time scaling}

These arguments carry over if we instead consider the transition as a function of time instead of system size. This is more directly relevant to ultracold atomic experiments, which currently seem to be limited more severely by coherence times than by system sizes.  We define a finite-time transition point $J_c(t)$ as that at which the typical spin has become entangled with many other degrees of freedom.  At time $t$ and small $J$, this happens if the distance $R$ to the nearest avalanche-inducing rare region satisfies $t \agt \exp(R/\zeta)$, where $\zeta \sim 1/(\log J)$. Thus, the probability of a typical spin entangling via an avalanche at time $t$ is 
\beq\label{critt}
P(t, J) \sim |\log t / \log J|^d \exp(-|\log^3 J|).
\eeq
This gives, to leading order, $\log J_c(t) \sim -(\log \log t)^{1/3}$. Also, analogous to the reasoning above, 
$\delta J / J_c(t) \sim (\log \log t)^{-2/3}$ and the transition is asymptotically well-defined (although the sharpening is extremely slow). 

This argument also suggests that, in the avalanche scenario, experiments with ultracold atoms should see an apparent MBL transition at an apparent critical coupling $J_c(t)$ that is only very weakly dependent on time.  This very weak drift of the transition point with time may be challenging to detect (Fig.~\ref{fig1}).

\subsection{Properties of the MBL and thermal phases}

We briefly discuss the low-frequency 
dynamics in this nonstandard scaling limit. 

First, we consider the MBL ``phase''; in this phase, in the frequency regime $\omega \ll J$, response is dominated by isolated local resonances (i.e., pairs of locally different configurations that hybridize). We count these resonances as follows~\cite{Gopalakrishnan15, mott1968}:  Given a frequency $\omega$, response at that frequency is dominated by transitions involving 
interactions between $n$ spins such that $J^n \simeq \omega$. The matrix elements of a generic local operator connecting the two resulting hybridized states are of order unity, and the number of such resonances is $z^{- \log \omega / \log J} \sim \omega^{-\log z/\log J}$ where $z$ is a number that depends on the dimensionality and lattice geometry. Thus a generic sample-averaged response function goes as $\chi(\omega) \sim \omega^{-\log z/\log J}$, so is present and only weakly dependent on frequency in the limit of large $L$  
~\footnote{In a typical sample, this response cuts off at some frequency $\omega_0$, set by the lowest-frequency resonance that occurs in that sample.  This is determined by the condition $(z J)^n L^d = 1$, which implies $\omega_0 \sim J^{-\log L^d / \log (zJ)} \sim L^{-d}$ (up to logarithms).}  This is nevertheless parametrically smaller than $J$, so a typical sample has a parametrically large window of low-frequency response. There are also contributions to the response due to rare thermal inclusions which are sub-critical in size, but these are subleading to these resonances.

At the transition $J_c(L)$, our scaling implies that a sample contains $O(1)$ supercritical inclusion cores.  Typical degrees of freedom relax by coupling to these cores. 
This coupling scales as $J^{a L}$, where $a$ is a constant.  Using the scaling of $J_c$ with $L$, we find that this typical relaxation time scales as $\log \tau \sim L (\log L)^{1/3}$.  This ``Thouless time'' of a sample thus grows slightly faster than exponentially with $L$, as a consequence of our scaling of $J_c(L)$. 

In the thermal phase, at large enough $L$ and $t$ so $J>J_c$, the system contains many thermal inclusions that are large enough to start avalanches.  The typical distance between inclusions is $R=R(J) \sim \exp(\text{constant} \times |\log^3 J|)$.  The time for a typical spin to get entangled with the nearest such inclusion is $\sim J^R$, after which that operator will spread by couplings between inclusions at some butterfly speed $v_B \sim R~J^{-R}$.

\section{Rapidly decaying power laws $\alpha > 2d$}\label{fastpower}

We turn next to rapidly decaying power laws, $1/r^\alpha$ with $\alpha > 2d$.  For these power laws, the MBL phase remains stable at weak interactions against few-spin resonances, so the channel by which it first thermalizes is an avalanche initiated by a rare supercritical thermal inclusion.  We now estimate the size and properties of such an inclusion.  As in the previous section, we assume we are at large enough $L$ and $t$ so that the transition is indeed set by when the avalanches get started.  At smaller $L$ or $t$, a Fock-space localization estimate of the crossover may be more appropriate.

\subsection{Constructing the critical inclusion}

The main difference between the short-range and power-law cases is that inclusions in the power-law case can be much less compact.  We can construct them as follows:  The inclusion has a ``core'' of closely-spaced resonant spins that has a bandwidth $J$, as in the short range case.  As the inclusion adds more spins, it becomes capable of absorbing spins that are increasingly far from its core.  For instance, if the inclusion contains $q$ spins, it can grow to $q + 1$ spins by incorporating any spin at a distance $R_q$ such that $1/R_q^{2\alpha} \agt 2^{-q}$.  Thus, $R_{q} \alt 2^{q/(2\alpha)}$.  The probability of there being a spin within this distance that is resonant with the core of the inclusion is $\sim R_{q}^d J$.  Once this probability reaches one, the inclusion has started an avalanche.
The critical inclusion size is thus an inclusion with $\ell$ spins, where
\beq\label{plinc}
\ell \simeq \frac{-2 \alpha \log_2 J}{d},
\eeq
with the farthest spin within the inclusion being at a distance 
$\sim R_{\ell} \sim J^{-1/d}$ from the core. 

\subsection{Location and width of the transition}

Given the critical inclusion size~\eqref{plinc} 
one can see that the probability of such an inclusion being present is  
\beq
P(J,L) \sim L^d \exp(-\alpha\times{\rm constant}\times\log^2 J)~.
\eeq
This gives $\log J_c(L) \sim -\sqrt{(1/\alpha) \log L}$. One can check immediately that this scaling corresponds to a transition that sharpens in the $L \rightarrow \infty$ limit as $(\log L)^{-1/2}$. 
Note also that the size of the critical inclusion will scale as $\log R_{\ell} \propto \sqrt{\log L}$, i.e., the critical inclusion, though spatially sparse, does grow slower than algebraically with system size. 

\subsection{Finite-time scaling}

We now turn from finite-size scaling to finite-time scaling. A typical spin couples to the nearest avalanche, which is a distance 
\beq\label{rj_pl}
R=R(J) \sim \exp(\alpha\times \text{constant} \times \log^2 J)
\eeq 
away, by a multi-spin 
process that is on shell to within $J$. Since $J \rightarrow 0$, it is optimal to use as few powers of $J$ as possible, by only going to second order.  To find a partner that is within $J$ in energy, we must typically go a distance $\sim J^{-1/d}$ from the core of the inclusion, so the rate at which a typical spin entangles with an inclusion at distance $R$ is $\sim J^{1 + 2\alpha/d}/{R^{2\alpha}}$.  However, $R$ increases faster than any power of $J$, so to leading order we obtain the length-time relationship $t(R) \sim R^{2\alpha}$. (Although we did not systematically optimize over all processes, $t(R)$ 
can not grow parametrically faster than $R^{2\alpha}$.)
Therefore the transition happens when $\log J \sim \log^{1/2} t$ and the $t$-driven transition has a width that scales down with $t$ as $\delta J/J_c(t) \sim \log^{-1/2} t$. 

\subsection{Properties of the localized and thermal phases}

At sufficiently low frequencies in the localized phase, one expects two-level systems to dominate response. For power-law interactions, these are primarily two-site resonances (as opposed to the many-body rearrangements that occur in the short-range case). At a frequency $\omega$, the dominant resonances are those at distance $(J/\omega)^{1/\alpha}$. The low-frequency response will therefore scale as $R(\omega) \sim \omega^{2 - d/\alpha}$. 
At the transition, the bottleneck for relaxation is the rate at which a typical spin couples to the supercritical inclusion (recall that near the transition there are typically $O(1)$ such inclusions per sample of size $L$).  To leading order, then, $\tau \sim L^{2\alpha}$. There will be multiplicative corrections due to the $L$-dependence of $J$; however, this dependence is slower than any power of $L$ and we neglect it.  In the thermal phase, the inclusions are spaced by $R(J)$ [Eq.~\eqref{rj_pl}], and the Golden-Rule timescale $t(R) \sim R^{2\alpha}/J$ governs the relaxation of typical spins. Operators spread via these inclusions with butterfly speed $v_B \sim [R(J)]^{-(2\alpha - 1)}$; {although spreading operators will have power-law tails, when $\alpha > 2d$ the width of the operator front, and therefore $v_B$, remain well-defined~\cite{fossfeig2015, cz2018}}.  

\section{One-dimensional systems}\label{1dweak}

In one dimension, the MBL transition in the conventional thermodynamic limit can be determined, not by rare regions that initiate avalanches, but by the localization properties of \emph{typical} regions.  When typical regions have a sufficiently short decay length $\zeta$ for spin-flip interactions, even large avalanches do not propagate far; when this decay length exceeds a critical value, a single avalanche can destabilize the entire system \cite{drh,ldrh}.  A transition of this sort (involving the spreading of avalanches) has been seen in approximate RG studies of the MBL transition that study very large systems~\cite{vha, pvp, dvp, thmdr, dumitrescu2018kosterlitz}.  {The regime being studied in all of these RG's is length scales where thermal inclusions large enough to start an avalanche are present.}  The critical exponents seen in these RG studies are all 
consistent with the CCFS bound on critical exponents in random systems~\cite{ccfs,clo}, with the most recent work supporting a Kosterlitz-Thouless-like transition~\cite{dumitrescu2018kosterlitz}.
However, exact diagonalization (ED) studies of the transition, which are necessarily confined to small system sizes, find very different apparent critical exponents that are inconsistent with CCFS~\cite{kbp, lla, ksh}.  This suggests that the transitions seen in the ED studies may be in a finite-size regime that is very different from the thermodynamic limit.

\begin{figure}[!t]
\begin{center}
\includegraphics[width = 0.4 \textwidth]{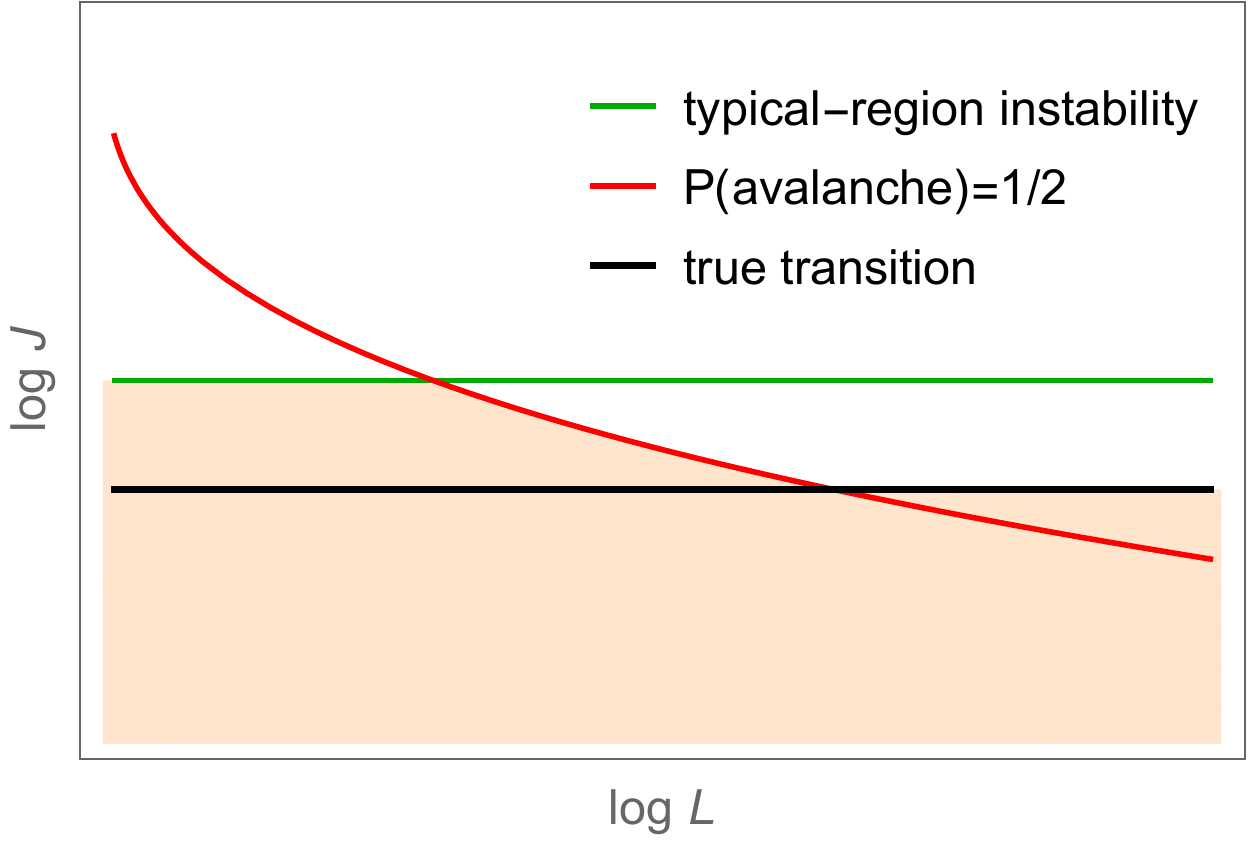}
\caption{Schematic illustration of the scenario discussed in the main text for MBL transitions $J_c(L)$ vs. system size $L$ in one dimensional systems with short-range interactions.  At the smallest sizes, avalanches are rare and the transition occurs when typical regions are unstable to the proliferation of resonances (green line).  At the largest sizes, the transition occurs when an avalanche is able to grow without bound (black line).  As one increases the system size, one goes through a regime of intermediate sizes where the bottleneck for thermalization is having a rare region that is large enough to start an avalanche that can grow (red curve).  In models with correlations in the disorder, the density of avalanches can be tuned separately from the typical localization length, allowing one in principle to slide the red curve horizontally (Sec.~\ref{disc}). {Further, the black and green lines can be shifted relative to one another, e.g., for interacting {particles,} by tuning the relative strength of hopping and interactions.}}
\label{1dspec}
\end{center}
\end{figure}

One scenario for this discrepancy is that the numerically observed transition in small systems occurs when typical regions develop resonances. In this scenario, avalanches are too rare to be relevant at the {numerically} accessible system sizes. Then, as one increases the system size, avalanches begin to become relevant and move the transition to lower interaction; however, for intermediate sizes, the bottleneck for a transition to occur in a given sample is whether that sample hosts a large enough locally thermalizing rare region to start an avalanche.  As the system grows, this avalanche-start ``transition'' drifts to increasingly small interactions, and for large enough systems takes place inside the MBL phase.  {It is only once the avalanche-start transition has moved to well below the true avalanche runaway transition that the system will display the asymptotic critical behavior of the thermodynamic limit.}  But this might not happen until lengths $L$ that greatly exceed those accessible to ED. 
This scenario is sketched in Fig.~\ref{1dspec}. {In the much-studied case of Heisenberg chains with random on-site fields in the $z$ direction, numerical evidence~\cite{varma2019, cappellaro} suggests that the localization length remains short enough to contain avalanches throughout most of the numerically observed localized phase, i.e., the typical-region instability and the avalanche instability occur at comparable} {interactions.}

One can also construct one-dimensional models in which the transition is instead bottlenecked by the density of inclusions out to the large-$L$ limit.  For instance, if the microscopic interactions themselves fall off exponentially with a slow enough decay constant, or for a weakly-interacting Anderson insulator with {strong hopping and thus} a relatively large localization length, starting an avalanche remains the bottleneck even for large system sizes. 

For the weakly-interacting {and weakly-localized one-dimensional} Anderson insulator, we can write out the Hamiltonian in the basis of localized single-particle eigenstates $\alpha$, as follows: 
\beq
H = \sum_\alpha \epsilon_\alpha n_\alpha + \sum_{\alpha\beta\gamma\delta} V_{\alpha\beta\gamma\delta} c^\dagger_\alpha c^\dagger_\beta c_\gamma c_\delta.
\eeq
Here, the interaction coefficients fall off exponentially with the distance between the centers of the orbitals involved. (Specifically, the typical falloff is exponential, but there is also an exponentially small probability of an $O(1)$ coupling.)  Although the interaction here is a four-site coupling rather than the two-site coupling considered earlier, we can construct rare regions analogous to the higher-dimensional short-range case.  The role of a spin flip is played here by a particle-hole excitation involving two overlapping localized orbitals.  The rare inclusion is a cluster of $\ell$ nearby such particle-hole excitations that all have the same energy within the interaction $|V|$.
These make up a narrow-bandwidth bath with bandwidth $\sim V$. 
The rest of the argument proceeds as in the $d > 1$ case (Sec.~\ref{higherd}), giving the same scaling of $J_c(L)$ [Eq.~\eqref{sravalanche}] that we previously found for short-range models.

\section{Discussion}\label{disc}

In this work we have argued that, even assuming the MBL transition occurs via an ``avalanche'' instability, there is a sense in which the transition remains sharp and well-defined in dimensions greater than one.  Defining this transition requires taking an unconventional thermodynamic limit, in which the interaction $J$ is reduced in proportion to the critical value $J_c(L)$ that goes to zero
as $L \rightarrow \infty$.  With this scaling, the transition is sharp in this unusual thermodynamic limit, since as $L \rightarrow \infty$ the probability that a given sample contains a supercritical inclusion becomes a step function of $(J - J_c(L))/J_c(L)$ (Fig.~1).  This transition has some features in common with other extreme value problems, such as the emergence of a sharp density-of-states edge in random matrices~\cite{majumdar}, which it would be interesting to explore further. 

Our estimates imply that in short-range models in $d > 1$, the drift of the apparent transition $J_c(t)$ with the observation time $t$ is extremely weak, 
 and may be effectively unobservable in many experiments.  Thus, in practice such experiments may see a constant, well-defined transition $J_c(t)$ that shows no discernible time dependence.  However, for models with rapidly decaying power law interactions, such as dipolar systems~\cite{ylg}, the time-dependence of $J_c(t)$ is enhanced and might more readily be observed.

A key assumption in this work was that disorder is uncorrelated and drawn from a smooth distribution; thus the same parameter controls the localization parameters of typical regions and the density of rare thermal inclusions.  However, if one works in a space of more general disorder models---e.g., correlated random potentials, or disorder distributions with sharp peaks at certain specific values---these two properties can be controlled separately.  The scaling of the MBL transition might differ in these more general models from what we have discussed above.  An extreme case is that of quasiperiodic potentials.  Whether avalanches can arise in localized quasiperiodic systems is not known at present, although the single-particle localized state is unstable to weak interactions when $\xi$ is of order unity~\cite{vznidarivc2018interaction}.  In one-dimensional quasiperiodic potentials the probability of finding a sequence of $n > 2$ adjacent sites that are degenerate to a given threshold is strictly zero.  Therefore avalanches, even if they do get started, must grow from seeds that look very different than those considered here.  At the opposite extreme, locally correlated disorder potentials can have an anomalously high density of supercritical inclusions, and thus should be able to exhibit the ``true'' MBL transition for short-range 1D systems for smaller size samples, perhaps even for those accessible to exact diagonalization. 

\begin{acknowledgments}
The authors are grateful to Vedika Khemani and Vadim Oganesyan for helpful discussions.  S. G. acknowledges support from NSF Grant No. DMR-1653271.  D. H. is supported in part by the DARPA DRINQS program.
\end{acknowledgments}

\bibliography{QPTFIMcriticality,library}

\begin{thebibliography}{53}
\expandafter\ifx\csname natexlab\endcsname\relax\def\natexlab#1{#1}\fi
\expandafter\ifx\csname bibnamefont\endcsname\relax
  \def\bibnamefont#1{#1}\fi
\expandafter\ifx\csname bibfnamefont\endcsname\relax
  \def\bibfnamefont#1{#1}\fi
\expandafter\ifx\csname citenamefont\endcsname\relax
  \def\citenamefont#1{#1}\fi
\expandafter\ifx\csname url\endcsname\relax
  \def\url#1{\texttt{#1}}\fi
\expandafter\ifx\csname urlprefix\endcsname\relax\def\urlprefix{URL }\fi
\providecommand{\bibinfo}[2]{#2}
\providecommand{\eprint}[2][]{\url{#2}}

\bibitem[{\citenamefont{Basko et~al.}(2006)\citenamefont{Basko, Aleiner, and
  Altshuler}}]{basko_metalinsulator_2006}
\bibinfo{author}{\bibfnamefont{D.}~\bibnamefont{Basko}},
  \bibinfo{author}{\bibfnamefont{I.}~\bibnamefont{Aleiner}}, \bibnamefont{and}
  \bibinfo{author}{\bibfnamefont{B.}~\bibnamefont{Altshuler}},
  \bibinfo{journal}{Ann. Phys. (N.Y.)} \textbf{\bibinfo{volume}{321}},
  \bibinfo{pages}{1126} (\bibinfo{year}{2006}),
  \urlprefix\url{http://www.sciencedirect.com/science/article/pii/S0003491605002630}.

\bibitem[{\citenamefont{Gornyi et~al.}(2005)\citenamefont{Gornyi, Mirlin, and
  Polyakov}}]{gornyi_interacting_2005}
\bibinfo{author}{\bibfnamefont{I.~V.} \bibnamefont{Gornyi}},
  \bibinfo{author}{\bibfnamefont{A.~D.} \bibnamefont{Mirlin}},
  \bibnamefont{and} \bibinfo{author}{\bibfnamefont{D.~G.}
  \bibnamefont{Polyakov}}, \bibinfo{journal}{Phys. Rev. Lett.}
  \textbf{\bibinfo{volume}{95}}, \bibinfo{pages}{206603}
  (\bibinfo{year}{2005}),
  \urlprefix\url{http://link.aps.org/doi/10.1103/PhysRevLett.95.206603}.

\bibitem[{\citenamefont{Oganesyan and
  Huse}(2007)}]{oganesyan_localization_2007}
\bibinfo{author}{\bibfnamefont{V.}~\bibnamefont{Oganesyan}} \bibnamefont{and}
  \bibinfo{author}{\bibfnamefont{D.~A.} \bibnamefont{Huse}},
  \bibinfo{journal}{Phys. Rev. B} \textbf{\bibinfo{volume}{75}},
  \bibinfo{pages}{155111} (\bibinfo{year}{2007}),
  \urlprefix\url{http://link.aps.org/doi/10.1103/PhysRevB.75.155111}.

\bibitem[{\citenamefont{{\v Z}nidari{\v c} et~al.}(2008)\citenamefont{{\v
  Z}nidari{\v c}, Prosen, and Prelov{\v s}ek}}]{znidaric_many-body_2008}
\bibinfo{author}{\bibfnamefont{M.}~\bibnamefont{{\v Z}nidari{\v c}}},
  \bibinfo{author}{\bibfnamefont{T.}~\bibnamefont{Prosen}}, \bibnamefont{and}
  \bibinfo{author}{\bibfnamefont{P.}~\bibnamefont{Prelov{\v s}ek}},
  \bibinfo{journal}{Phys. Rev. B} \textbf{\bibinfo{volume}{77}},
  \bibinfo{pages}{064426} (\bibinfo{year}{2008}),
  \urlprefix\url{http://link.aps.org/doi/10.1103/PhysRevB.77.064426}.

\bibitem[{\citenamefont{Nandkishore and Huse}(2015)}]{nandkishore_mbl_2015}
\bibinfo{author}{\bibfnamefont{R.}~\bibnamefont{Nandkishore}} \bibnamefont{and}
  \bibinfo{author}{\bibfnamefont{D.~A.} \bibnamefont{Huse}},
  \bibinfo{journal}{Annu. Rev. Condens. Matter} \textbf{\bibinfo{volume}{6}},
  \bibinfo{pages}{15} (\bibinfo{year}{2015}),
  \urlprefix\url{http://dx.doi.org/10.1146/annurev-conmatphys-031214-014726}.

\bibitem[{\citenamefont{Serbyn et~al.}(2013{\natexlab{a}})\citenamefont{Serbyn,
  Papi{\'c}, and Abanin}}]{serbyn_universal_2013}
\bibinfo{author}{\bibfnamefont{M.}~\bibnamefont{Serbyn}},
  \bibinfo{author}{\bibfnamefont{Z.}~\bibnamefont{Papi{\'c}}},
  \bibnamefont{and} \bibinfo{author}{\bibfnamefont{D.~A.}
  \bibnamefont{Abanin}}, \bibinfo{journal}{Phys. Rev. Lett.}
  \textbf{\bibinfo{volume}{110}}, \bibinfo{pages}{260601}
  (\bibinfo{year}{2013}{\natexlab{a}}).

\bibitem[{\citenamefont{Serbyn et~al.}(2013{\natexlab{b}})\citenamefont{Serbyn,
  Papi{\'c}, and Abanin}}]{serbyn_local_2013}
\bibinfo{author}{\bibfnamefont{M.}~\bibnamefont{Serbyn}},
  \bibinfo{author}{\bibfnamefont{Z.}~\bibnamefont{Papi{\'c}}},
  \bibnamefont{and} \bibinfo{author}{\bibfnamefont{D.~A.}
  \bibnamefont{Abanin}}, \bibinfo{journal}{Phys. Rev. Lett.}
  \textbf{\bibinfo{volume}{111}}, \bibinfo{pages}{127201}
  (\bibinfo{year}{2013}{\natexlab{b}}).

\bibitem[{\citenamefont{Huse et~al.}(2014)\citenamefont{Huse, Nandkishore, and
  Oganesyan}}]{hno}
\bibinfo{author}{\bibfnamefont{D.~A.} \bibnamefont{Huse}},
  \bibinfo{author}{\bibfnamefont{R.}~\bibnamefont{Nandkishore}},
  \bibnamefont{and}
  \bibinfo{author}{\bibfnamefont{V.}~\bibnamefont{Oganesyan}},
  \bibinfo{journal}{Phys. Rev. B} \textbf{\bibinfo{volume}{90}},
  \bibinfo{pages}{174202} (\bibinfo{year}{2014}),
  \urlprefix\url{https://link.aps.org/doi/10.1103/PhysRevB.90.174202}.

\bibitem[{\citenamefont{Schreiber et~al.}(2015)\citenamefont{Schreiber,
  Hodgman, Bordia, L\"uschen, Fischer, Vosk, Altman, Schneider, and
  Bloch}}]{Schreiber15}
\bibinfo{author}{\bibfnamefont{M.}~\bibnamefont{Schreiber}},
  \bibinfo{author}{\bibfnamefont{S.~S.} \bibnamefont{Hodgman}},
  \bibinfo{author}{\bibfnamefont{P.}~\bibnamefont{Bordia}},
  \bibinfo{author}{\bibfnamefont{H.~P.} \bibnamefont{L\"uschen}},
  \bibinfo{author}{\bibfnamefont{M.~H.} \bibnamefont{Fischer}},
  \bibinfo{author}{\bibfnamefont{R.}~\bibnamefont{Vosk}},
  \bibinfo{author}{\bibfnamefont{E.}~\bibnamefont{Altman}},
  \bibinfo{author}{\bibfnamefont{U.}~\bibnamefont{Schneider}},
  \bibnamefont{and} \bibinfo{author}{\bibfnamefont{I.}~\bibnamefont{Bloch}},
  \bibinfo{journal}{Science} \textbf{\bibinfo{volume}{349}},
  \bibinfo{pages}{842} (\bibinfo{year}{2015}),
  \urlprefix\url{http://www.sciencemag.org/content/349/6250/842.abstract}.

\bibitem[{\citenamefont{Imbrie}(2016{\natexlab{a}})}]{jzi}
\bibinfo{author}{\bibfnamefont{J.~Z.} \bibnamefont{Imbrie}},
  \bibinfo{journal}{Journal of Statistical Physics}
  \textbf{\bibinfo{volume}{163}}, \bibinfo{pages}{998}
  (\bibinfo{year}{2016}{\natexlab{a}}).

\bibitem[{\citenamefont{Imbrie}(2016{\natexlab{b}})}]{jzi2}
\bibinfo{author}{\bibfnamefont{J.~Z.} \bibnamefont{Imbrie}},
  \bibinfo{journal}{Phys. Rev. Lett.} \textbf{\bibinfo{volume}{117}},
  \bibinfo{pages}{027201} (\bibinfo{year}{2016}{\natexlab{b}}),
  \urlprefix\url{https://link.aps.org/doi/10.1103/PhysRevLett.117.027201}.

\bibitem[{\citenamefont{Bauer and Nayak}(2013)}]{bauer_area_2013}
\bibinfo{author}{\bibfnamefont{B.}~\bibnamefont{Bauer}} \bibnamefont{and}
  \bibinfo{author}{\bibfnamefont{C.}~\bibnamefont{Nayak}}, \bibinfo{journal}{J.
  Stat. Mech.} \textbf{\bibinfo{volume}{2013}}, \bibinfo{pages}{P09005}
  (\bibinfo{year}{2013}).

\bibitem[{\citenamefont{Choi et~al.}(2016)\citenamefont{Choi, Hild, Zeiher,
  Schau{\ss}, Rubio-Abadal, Yefsah, Khemani, Huse, Bloch, and Gross}}]{Hild16}
\bibinfo{author}{\bibfnamefont{J.-y.} \bibnamefont{Choi}},
  \bibinfo{author}{\bibfnamefont{S.}~\bibnamefont{Hild}},
  \bibinfo{author}{\bibfnamefont{J.}~\bibnamefont{Zeiher}},
  \bibinfo{author}{\bibfnamefont{P.}~\bibnamefont{Schau{\ss}}},
  \bibinfo{author}{\bibfnamefont{A.}~\bibnamefont{Rubio-Abadal}},
  \bibinfo{author}{\bibfnamefont{T.}~\bibnamefont{Yefsah}},
  \bibinfo{author}{\bibfnamefont{V.}~\bibnamefont{Khemani}},
  \bibinfo{author}{\bibfnamefont{D.~A.} \bibnamefont{Huse}},
  \bibinfo{author}{\bibfnamefont{I.}~\bibnamefont{Bloch}}, \bibnamefont{and}
  \bibinfo{author}{\bibfnamefont{C.}~\bibnamefont{Gross}},
  \bibinfo{journal}{Science} \textbf{\bibinfo{volume}{352}},
  \bibinfo{pages}{1547} (\bibinfo{year}{2016}), ISSN \bibinfo{issn}{0036-8075},
  \urlprefix\url{http://science.sciencemag.org/content/352/6293/1547}.

\bibitem[{\citenamefont{Bordia et~al.}(2016)\citenamefont{Bordia, L\"uschen,
  Hodgman, Schreiber, Bloch, and Schneider}}]{Bordia16}
\bibinfo{author}{\bibfnamefont{P.}~\bibnamefont{Bordia}},
  \bibinfo{author}{\bibfnamefont{H.~P.} \bibnamefont{L\"uschen}},
  \bibinfo{author}{\bibfnamefont{S.~S.} \bibnamefont{Hodgman}},
  \bibinfo{author}{\bibfnamefont{M.}~\bibnamefont{Schreiber}},
  \bibinfo{author}{\bibfnamefont{I.}~\bibnamefont{Bloch}}, \bibnamefont{and}
  \bibinfo{author}{\bibfnamefont{U.}~\bibnamefont{Schneider}},
  \bibinfo{journal}{Phys. Rev. Lett.} \textbf{\bibinfo{volume}{116}},
  \bibinfo{pages}{140401} (\bibinfo{year}{2016}),
  \urlprefix\url{http://link.aps.org/doi/10.1103/PhysRevLett.116.140401}.

\bibitem[{\citenamefont{Bordia et~al.}(2017)\citenamefont{Bordia, L\"uschen,
  Scherg, Gopalakrishnan, Knap, Schneider, and Bloch}}]{bordia2d17}
\bibinfo{author}{\bibfnamefont{P.}~\bibnamefont{Bordia}},
  \bibinfo{author}{\bibfnamefont{H.}~\bibnamefont{L\"uschen}},
  \bibinfo{author}{\bibfnamefont{S.}~\bibnamefont{Scherg}},
  \bibinfo{author}{\bibfnamefont{S.}~\bibnamefont{Gopalakrishnan}},
  \bibinfo{author}{\bibfnamefont{M.}~\bibnamefont{Knap}},
  \bibinfo{author}{\bibfnamefont{U.}~\bibnamefont{Schneider}},
  \bibnamefont{and} \bibinfo{author}{\bibfnamefont{I.}~\bibnamefont{Bloch}},
  \bibinfo{journal}{Phys. Rev. X} \textbf{\bibinfo{volume}{7}},
  \bibinfo{pages}{041047} (\bibinfo{year}{2017}),
  \urlprefix\url{https://link.aps.org/doi/10.1103/PhysRevX.7.041047}.

\bibitem[{\citenamefont{Wahl et~al.}(2017)\citenamefont{Wahl, Pal, and
  Simon}}]{wahl2017signatures}
\bibinfo{author}{\bibfnamefont{T.~B.} \bibnamefont{Wahl}},
  \bibinfo{author}{\bibfnamefont{A.}~\bibnamefont{Pal}}, \bibnamefont{and}
  \bibinfo{author}{\bibfnamefont{S.~H.} \bibnamefont{Simon}},
  \bibinfo{journal}{arXiv preprint arXiv:1711.02678}  (\bibinfo{year}{2017}).

\bibitem[{\citenamefont{Burin et~al.}(1998)\citenamefont{Burin, Kagan,
  Maksimov, and Polishchuk}}]{burin1998}
\bibinfo{author}{\bibfnamefont{A.~L.} \bibnamefont{Burin}},
  \bibinfo{author}{\bibfnamefont{Y.}~\bibnamefont{Kagan}},
  \bibinfo{author}{\bibfnamefont{L.~A.} \bibnamefont{Maksimov}},
  \bibnamefont{and} \bibinfo{author}{\bibfnamefont{I.~Y.}
  \bibnamefont{Polishchuk}}, \bibinfo{journal}{Phys. Rev. Lett.}
  \textbf{\bibinfo{volume}{80}}, \bibinfo{pages}{2945} (\bibinfo{year}{1998}),
  \urlprefix\url{https://link.aps.org/doi/10.1103/PhysRevLett.80.2945}.

\bibitem[{\citenamefont{Burin}(2006)}]{burin2006}
\bibinfo{author}{\bibfnamefont{A.~L.} \bibnamefont{Burin}},
  \bibinfo{journal}{arXiv preprint cond-mat/0611387}  (\bibinfo{year}{2006}).

\bibitem[{\citenamefont{Yao et~al.}(2014)\citenamefont{Yao, Laumann,
  Gopalakrishnan, Knap, M\"uller, Demler, and Lukin}}]{ylg}
\bibinfo{author}{\bibfnamefont{N.~Y.} \bibnamefont{Yao}},
  \bibinfo{author}{\bibfnamefont{C.~R.} \bibnamefont{Laumann}},
  \bibinfo{author}{\bibfnamefont{S.}~\bibnamefont{Gopalakrishnan}},
  \bibinfo{author}{\bibfnamefont{M.}~\bibnamefont{Knap}},
  \bibinfo{author}{\bibfnamefont{M.}~\bibnamefont{M\"uller}},
  \bibinfo{author}{\bibfnamefont{E.~A.} \bibnamefont{Demler}},
  \bibnamefont{and} \bibinfo{author}{\bibfnamefont{M.~D.} \bibnamefont{Lukin}},
  \bibinfo{journal}{Phys. Rev. Lett.} \textbf{\bibinfo{volume}{113}},
  \bibinfo{pages}{243002} (\bibinfo{year}{2014}),
  \urlprefix\url{https://link.aps.org/doi/10.1103/PhysRevLett.113.243002}.

\bibitem[{\citenamefont{Burin}(2015{\natexlab{a}})}]{burin2015}
\bibinfo{author}{\bibfnamefont{A.~L.} \bibnamefont{Burin}},
  \bibinfo{journal}{Phys. Rev. B} \textbf{\bibinfo{volume}{92}},
  \bibinfo{pages}{104428} (\bibinfo{year}{2015}{\natexlab{a}}),
  \urlprefix\url{https://link.aps.org/doi/10.1103/PhysRevB.92.104428}.

\bibitem[{\citenamefont{Burin}(2015{\natexlab{b}})}]{burin2015b}
\bibinfo{author}{\bibfnamefont{A.~L.} \bibnamefont{Burin}},
  \bibinfo{journal}{Phys. Rev. B} \textbf{\bibinfo{volume}{91}},
  \bibinfo{pages}{094202} (\bibinfo{year}{2015}{\natexlab{b}}),
  \urlprefix\url{https://link.aps.org/doi/10.1103/PhysRevB.91.094202}.

\bibitem[{\citenamefont{Gornyi et~al.}(2017)\citenamefont{Gornyi, Mirlin,
  Polyakov, and Burin}}]{gornyi2017spectral}
\bibinfo{author}{\bibfnamefont{I.}~\bibnamefont{Gornyi}},
  \bibinfo{author}{\bibfnamefont{A.}~\bibnamefont{Mirlin}},
  \bibinfo{author}{\bibfnamefont{D.}~\bibnamefont{Polyakov}}, \bibnamefont{and}
  \bibinfo{author}{\bibfnamefont{A.}~\bibnamefont{Burin}},
  \bibinfo{journal}{Annalen der Physik} \textbf{\bibinfo{volume}{529}},
  \bibinfo{pages}{1600360} (\bibinfo{year}{2017}).

\bibitem[{\citenamefont{Gutman et~al.}(2016)\citenamefont{Gutman, Protopopov,
  Burin, Gornyi, Santos, and Mirlin}}]{PhysRevB.93.245427}
\bibinfo{author}{\bibfnamefont{D.~B.} \bibnamefont{Gutman}},
  \bibinfo{author}{\bibfnamefont{I.~V.} \bibnamefont{Protopopov}},
  \bibinfo{author}{\bibfnamefont{A.~L.} \bibnamefont{Burin}},
  \bibinfo{author}{\bibfnamefont{I.~V.} \bibnamefont{Gornyi}},
  \bibinfo{author}{\bibfnamefont{R.~A.} \bibnamefont{Santos}},
  \bibnamefont{and} \bibinfo{author}{\bibfnamefont{A.~D.}
  \bibnamefont{Mirlin}}, \bibinfo{journal}{Phys. Rev. B}
  \textbf{\bibinfo{volume}{93}}, \bibinfo{pages}{245427}
  (\bibinfo{year}{2016}),
  \urlprefix\url{https://link.aps.org/doi/10.1103/PhysRevB.93.245427}.

\bibitem[{\citenamefont{Tikhonov and Mirlin}(2018)}]{tikhonov2018}
\bibinfo{author}{\bibfnamefont{K.~S.} \bibnamefont{Tikhonov}} \bibnamefont{and}
  \bibinfo{author}{\bibfnamefont{A.~D.} \bibnamefont{Mirlin}},
  \bibinfo{journal}{Phys. Rev. B} \textbf{\bibinfo{volume}{97}},
  \bibinfo{pages}{214205} (\bibinfo{year}{2018}),
  \urlprefix\url{https://link.aps.org/doi/10.1103/PhysRevB.97.214205}.

\bibitem[{\citenamefont{Iyer et~al.}(2013)\citenamefont{Iyer, Oganesyan,
  Refael, and Huse}}]{iyer_many-body_2013}
\bibinfo{author}{\bibfnamefont{S.}~\bibnamefont{Iyer}},
  \bibinfo{author}{\bibfnamefont{V.}~\bibnamefont{Oganesyan}},
  \bibinfo{author}{\bibfnamefont{G.}~\bibnamefont{Refael}}, \bibnamefont{and}
  \bibinfo{author}{\bibfnamefont{D.~A.} \bibnamefont{Huse}},
  \bibinfo{journal}{Phys. Rev. B} \textbf{\bibinfo{volume}{87}},
  \bibinfo{pages}{134202} (\bibinfo{year}{2013}),
  \urlprefix\url{http://link.aps.org/doi/10.1103/PhysRevB.87.134202}.

\bibitem[{\citenamefont{L\"uschen et~al.}(2016)\citenamefont{L\"uschen, Bordia,
  Scherg, Alet, Altman, Schneider, and Bloch}}]{lueschencrit16}
\bibinfo{author}{\bibfnamefont{H.~P.} \bibnamefont{L\"uschen}},
  \bibinfo{author}{\bibfnamefont{P.}~\bibnamefont{Bordia}},
  \bibinfo{author}{\bibfnamefont{S.}~\bibnamefont{Scherg}},
  \bibinfo{author}{\bibfnamefont{F.}~\bibnamefont{Alet}},
  \bibinfo{author}{\bibfnamefont{E.}~\bibnamefont{Altman}},
  \bibinfo{author}{\bibfnamefont{U.}~\bibnamefont{Schneider}},
  \bibnamefont{and} \bibinfo{author}{\bibfnamefont{I.}~\bibnamefont{Bloch}}
  (\bibinfo{year}{2016}), \eprint{arXiv:1612.07173}.

\bibitem[{\citenamefont{Khemani et~al.}(2017)\citenamefont{Khemani, Sheng, and
  Huse}}]{ksh}
\bibinfo{author}{\bibfnamefont{V.}~\bibnamefont{Khemani}},
  \bibinfo{author}{\bibfnamefont{D.~N.} \bibnamefont{Sheng}}, \bibnamefont{and}
  \bibinfo{author}{\bibfnamefont{D.~A.} \bibnamefont{Huse}},
  \bibinfo{journal}{Phys. Rev. Lett.} \textbf{\bibinfo{volume}{119}},
  \bibinfo{pages}{075702} (\bibinfo{year}{2017}),
  \urlprefix\url{https://link.aps.org/doi/10.1103/PhysRevLett.119.075702}.

\bibitem[{\citenamefont{Setiawan et~al.}(2017)\citenamefont{Setiawan, Deng, and
  Pixley}}]{sdp}
\bibinfo{author}{\bibfnamefont{F.}~\bibnamefont{Setiawan}},
  \bibinfo{author}{\bibfnamefont{D.-L.} \bibnamefont{Deng}}, \bibnamefont{and}
  \bibinfo{author}{\bibfnamefont{J.~H.} \bibnamefont{Pixley}},
  \bibinfo{journal}{Phys. Rev. B} \textbf{\bibinfo{volume}{96}},
  \bibinfo{pages}{104205} (\bibinfo{year}{2017}),
  \urlprefix\url{https://link.aps.org/doi/10.1103/PhysRevB.96.104205}.

\bibitem[{\citenamefont{{\v{Z}}nidari{\v{c}} and
  Ljubotina}(2018)}]{vznidarivc2018interaction}
\bibinfo{author}{\bibfnamefont{M.}~\bibnamefont{{\v{Z}}nidari{\v{c}}}}
  \bibnamefont{and}
  \bibinfo{author}{\bibfnamefont{M.}~\bibnamefont{Ljubotina}},
  \bibinfo{journal}{Proceedings of the National Academy of Sciences}
  \textbf{\bibinfo{volume}{115}}, \bibinfo{pages}{4595} (\bibinfo{year}{2018}).

\bibitem[{\citenamefont{De~Roeck and Huveneers}(2017)}]{drh}
\bibinfo{author}{\bibfnamefont{W.}~\bibnamefont{De~Roeck}} \bibnamefont{and}
  \bibinfo{author}{\bibfnamefont{F.~m.~c.} \bibnamefont{Huveneers}},
  \bibinfo{journal}{Phys. Rev. B} \textbf{\bibinfo{volume}{95}},
  \bibinfo{pages}{155129} (\bibinfo{year}{2017}),
  \urlprefix\url{https://link.aps.org/doi/10.1103/PhysRevB.95.155129}.

\bibitem[{\citenamefont{Luitz et~al.}(2017)\citenamefont{Luitz, Huveneers, and
  De~Roeck}}]{ldrh}
\bibinfo{author}{\bibfnamefont{D.~J.} \bibnamefont{Luitz}},
  \bibinfo{author}{\bibfnamefont{F.~m.~c.} \bibnamefont{Huveneers}},
  \bibnamefont{and} \bibinfo{author}{\bibfnamefont{W.}~\bibnamefont{De~Roeck}},
  \bibinfo{journal}{Phys. Rev. Lett.} \textbf{\bibinfo{volume}{119}},
  \bibinfo{pages}{150602} (\bibinfo{year}{2017}),
  \urlprefix\url{https://link.aps.org/doi/10.1103/PhysRevLett.119.150602}.

\bibitem[{\citenamefont{Ponte et~al.}(2017)\citenamefont{Ponte, Laumann, Huse,
  and Chandran}}]{plhc}
\bibinfo{author}{\bibfnamefont{P.}~\bibnamefont{Ponte}},
  \bibinfo{author}{\bibfnamefont{C.}~\bibnamefont{Laumann}},
  \bibinfo{author}{\bibfnamefont{D.~A.} \bibnamefont{Huse}}, \bibnamefont{and}
  \bibinfo{author}{\bibfnamefont{A.}~\bibnamefont{Chandran}},
  \bibinfo{journal}{Phil. Trans. R. Soc. A} \textbf{\bibinfo{volume}{375}},
  \bibinfo{pages}{20160428} (\bibinfo{year}{2017}).

\bibitem[{\citenamefont{Ponte et~al.}(2015)\citenamefont{Ponte,
  Papi\ifmmode~\acute{c}\else \'{c}\fi{}, Huveneers, and Abanin}}]{Ponte15}
\bibinfo{author}{\bibfnamefont{P.}~\bibnamefont{Ponte}},
  \bibinfo{author}{\bibfnamefont{Z.}~\bibnamefont{Papi\ifmmode~\acute{c}\else
  \'{c}\fi{}}}, \bibinfo{author}{\bibfnamefont{F.~m.~c.}
  \bibnamefont{Huveneers}}, \bibnamefont{and}
  \bibinfo{author}{\bibfnamefont{D.~A.} \bibnamefont{Abanin}},
  \bibinfo{journal}{Phys. Rev. Lett.} \textbf{\bibinfo{volume}{114}},
  \bibinfo{pages}{140401} (\bibinfo{year}{2015}),
  \urlprefix\url{http://link.aps.org/doi/10.1103/PhysRevLett.114.140401}.

\bibitem[{\citenamefont{Abanin et~al.}(2014)\citenamefont{Abanin, De~Roeck, and
  Huveneers}}]{abanin_theory_2014}
\bibinfo{author}{\bibfnamefont{D.}~\bibnamefont{Abanin}},
  \bibinfo{author}{\bibfnamefont{W.}~\bibnamefont{De~Roeck}}, \bibnamefont{and}
  \bibinfo{author}{\bibfnamefont{F.}~\bibnamefont{Huveneers}},
  \bibinfo{journal}{{arXiv}:1412.4752}  (\bibinfo{year}{2014}),
  \urlprefix\url{http://arxiv.org/abs/1412.4752}.

\bibitem[{\citenamefont{Pekker et~al.}(2017)\citenamefont{Pekker, Clark,
  Oganesyan, and Refael}}]{pcor}
\bibinfo{author}{\bibfnamefont{D.}~\bibnamefont{Pekker}},
  \bibinfo{author}{\bibfnamefont{B.~K.} \bibnamefont{Clark}},
  \bibinfo{author}{\bibfnamefont{V.}~\bibnamefont{Oganesyan}},
  \bibnamefont{and} \bibinfo{author}{\bibfnamefont{G.}~\bibnamefont{Refael}},
  \bibinfo{journal}{Phys. Rev. Lett.} \textbf{\bibinfo{volume}{119}},
  \bibinfo{pages}{075701} (\bibinfo{year}{2017}),
  \urlprefix\url{https://link.aps.org/doi/10.1103/PhysRevLett.119.075701}.

\bibitem[{\citenamefont{Varma et~al.}(2019)}]{varma2019}
\bibinfo{author}{\bibfnamefont{V.~K.} \bibnamefont{Varma}}
  \bibnamefont{et~al.}, \bibinfo{journal}{arXiv:1901.02902}
  (\bibinfo{year}{2019}).

\bibitem[{\citenamefont{Ros et~al.}(2015)\citenamefont{Ros, M{\"u}ller, and
  Scardicchio}}]{ros2015integrals}
\bibinfo{author}{\bibfnamefont{V.}~\bibnamefont{Ros}},
  \bibinfo{author}{\bibfnamefont{M.}~\bibnamefont{M{\"u}ller}},
  \bibnamefont{and}
  \bibinfo{author}{\bibfnamefont{A.}~\bibnamefont{Scardicchio}},
  \bibinfo{journal}{Nuclear Physics B} \textbf{\bibinfo{volume}{891}},
  \bibinfo{pages}{420} (\bibinfo{year}{2015}).

\bibitem[{\citenamefont{Gopalakrishnan and
  Nandkishore}(2014)}]{gopalakrishnan2014mean}
\bibinfo{author}{\bibfnamefont{S.}~\bibnamefont{Gopalakrishnan}}
  \bibnamefont{and}
  \bibinfo{author}{\bibfnamefont{R.}~\bibnamefont{Nandkishore}},
  \bibinfo{journal}{Phys. Rev. B} \textbf{\bibinfo{volume}{90}},
  \bibinfo{pages}{224203} (\bibinfo{year}{2014}).

\bibitem[{\citenamefont{Gopalakrishnan
  et~al.}(2015)\citenamefont{Gopalakrishnan, M\"uller, Khemani, Knap, Demler,
  and Huse}}]{Gopalakrishnan15}
\bibinfo{author}{\bibfnamefont{S.}~\bibnamefont{Gopalakrishnan}},
  \bibinfo{author}{\bibfnamefont{M.}~\bibnamefont{M\"uller}},
  \bibinfo{author}{\bibfnamefont{V.}~\bibnamefont{Khemani}},
  \bibinfo{author}{\bibfnamefont{M.}~\bibnamefont{Knap}},
  \bibinfo{author}{\bibfnamefont{E.}~\bibnamefont{Demler}}, \bibnamefont{and}
  \bibinfo{author}{\bibfnamefont{D.~A.} \bibnamefont{Huse}},
  \bibinfo{journal}{Phys. Rev. B} \textbf{\bibinfo{volume}{92}},
  \bibinfo{pages}{104202} (\bibinfo{year}{2015}),
  \urlprefix\url{http://link.aps.org/doi/10.1103/PhysRevB.92.104202}.

\bibitem[{\citenamefont{Mott}(1968)}]{mott1968}
\bibinfo{author}{\bibfnamefont{N.~F.} \bibnamefont{Mott}},
  \bibinfo{journal}{Phil. Mag.} \textbf{\bibinfo{volume}{17}},
  \bibinfo{pages}{1259} (\bibinfo{year}{1968}).

\bibitem[{\citenamefont{Foss-Feig et~al.}(2015)\citenamefont{Foss-Feig, Gong,
  Clark, and Gorshkov}}]{fossfeig2015}
\bibinfo{author}{\bibfnamefont{M.}~\bibnamefont{Foss-Feig}},
  \bibinfo{author}{\bibfnamefont{Z.-X.} \bibnamefont{Gong}},
  \bibinfo{author}{\bibfnamefont{C.~W.} \bibnamefont{Clark}}, \bibnamefont{and}
  \bibinfo{author}{\bibfnamefont{A.~V.} \bibnamefont{Gorshkov}},
  \bibinfo{journal}{Phys. Rev. Lett.} \textbf{\bibinfo{volume}{114}},
  \bibinfo{pages}{157201} (\bibinfo{year}{2015}),
  \urlprefix\url{https://link.aps.org/doi/10.1103/PhysRevLett.114.157201}.

\bibitem[{\citenamefont{Chen and Zhou}(2018)}]{cz2018}
\bibinfo{author}{\bibfnamefont{X.}~\bibnamefont{Chen}} \bibnamefont{and}
  \bibinfo{author}{\bibfnamefont{T.}~\bibnamefont{Zhou}},
  \bibinfo{journal}{arXiv preprint arXiv:1808.09812}  (\bibinfo{year}{2018}).

\bibitem[{\citenamefont{Vosk et~al.}(2015)\citenamefont{Vosk, Huse, and
  Altman}}]{vha}
\bibinfo{author}{\bibfnamefont{R.}~\bibnamefont{Vosk}},
  \bibinfo{author}{\bibfnamefont{D.~A.} \bibnamefont{Huse}}, \bibnamefont{and}
  \bibinfo{author}{\bibfnamefont{E.}~\bibnamefont{Altman}},
  \bibinfo{journal}{Phys. Rev. X} \textbf{\bibinfo{volume}{5}},
  \bibinfo{pages}{031032} (\bibinfo{year}{2015}),
  \urlprefix\url{https://link.aps.org/doi/10.1103/PhysRevX.5.031032}.

\bibitem[{\citenamefont{Potter et~al.}(2015)\citenamefont{Potter, Vasseur, and
  Parameswaran}}]{pvp}
\bibinfo{author}{\bibfnamefont{A.~C.} \bibnamefont{Potter}},
  \bibinfo{author}{\bibfnamefont{R.}~\bibnamefont{Vasseur}}, \bibnamefont{and}
  \bibinfo{author}{\bibfnamefont{S.~A.} \bibnamefont{Parameswaran}},
  \bibinfo{journal}{Phys. Rev. X} \textbf{\bibinfo{volume}{5}},
  \bibinfo{pages}{031033} (\bibinfo{year}{2015}),
  \urlprefix\url{https://link.aps.org/doi/10.1103/PhysRevX.5.031033}.

\bibitem[{\citenamefont{Dumitrescu et~al.}(2017)\citenamefont{Dumitrescu,
  Vasseur, and Potter}}]{dvp}
\bibinfo{author}{\bibfnamefont{P.~T.} \bibnamefont{Dumitrescu}},
  \bibinfo{author}{\bibfnamefont{R.}~\bibnamefont{Vasseur}}, \bibnamefont{and}
  \bibinfo{author}{\bibfnamefont{A.~C.} \bibnamefont{Potter}},
  \bibinfo{journal}{Phys. Rev. Lett.} \textbf{\bibinfo{volume}{119}},
  \bibinfo{pages}{110604} (\bibinfo{year}{2017}),
  \urlprefix\url{https://link.aps.org/doi/10.1103/PhysRevLett.119.110604}.

\bibitem[{\citenamefont{Thiery et~al.}(2018)\citenamefont{Thiery, Huveneers,
  M\"uller, and De~Roeck}}]{thmdr}
\bibinfo{author}{\bibfnamefont{T.}~\bibnamefont{Thiery}},
  \bibinfo{author}{\bibfnamefont{F.~m.~c.} \bibnamefont{Huveneers}},
  \bibinfo{author}{\bibfnamefont{M.}~\bibnamefont{M\"uller}}, \bibnamefont{and}
  \bibinfo{author}{\bibfnamefont{W.}~\bibnamefont{De~Roeck}},
  \bibinfo{journal}{Phys. Rev. Lett.} \textbf{\bibinfo{volume}{121}},
  \bibinfo{pages}{140601} (\bibinfo{year}{2018}),
  \urlprefix\url{https://link.aps.org/doi/10.1103/PhysRevLett.121.140601}.

\bibitem[{\citenamefont{Dumitrescu et~al.}(2018)\citenamefont{Dumitrescu,
  Parameswaran, Goremykina, Serbyn, and Vasseur}}]{dumitrescu2018kosterlitz}
\bibinfo{author}{\bibfnamefont{P.~T.} \bibnamefont{Dumitrescu}},
  \bibinfo{author}{\bibfnamefont{S.~A.} \bibnamefont{Parameswaran}},
  \bibinfo{author}{\bibfnamefont{A.}~\bibnamefont{Goremykina}},
  \bibinfo{author}{\bibfnamefont{M.}~\bibnamefont{Serbyn}}, \bibnamefont{and}
  \bibinfo{author}{\bibfnamefont{R.}~\bibnamefont{Vasseur}},
  \bibinfo{journal}{arXiv preprint arXiv:1811.03103}  (\bibinfo{year}{2018}).

\bibitem[{\citenamefont{Chayes et~al.}(1986)\citenamefont{Chayes, Chayes,
  Fisher, and Spencer}}]{ccfs}
\bibinfo{author}{\bibfnamefont{J.~T.} \bibnamefont{Chayes}},
  \bibinfo{author}{\bibfnamefont{L.}~\bibnamefont{Chayes}},
  \bibinfo{author}{\bibfnamefont{D.~S.} \bibnamefont{Fisher}},
  \bibnamefont{and} \bibinfo{author}{\bibfnamefont{T.}~\bibnamefont{Spencer}},
  \bibinfo{journal}{Phys. Rev. Lett.} \textbf{\bibinfo{volume}{57}},
  \bibinfo{pages}{2999} (\bibinfo{year}{1986}),
  \urlprefix\url{https://link.aps.org/doi/10.1103/PhysRevLett.57.2999}.

\bibitem[{\citenamefont{Chandran et~al.}(2015)\citenamefont{Chandran, Laumann,
  and Oganesyan}}]{clo}
\bibinfo{author}{\bibfnamefont{A.}~\bibnamefont{Chandran}},
  \bibinfo{author}{\bibfnamefont{C.~R.} \bibnamefont{Laumann}},
  \bibnamefont{and}
  \bibinfo{author}{\bibfnamefont{V.}~\bibnamefont{Oganesyan}},
  \bibinfo{journal}{arXiv preprint arXiv:1509.04285}  (\bibinfo{year}{2015}).

\bibitem[{\citenamefont{Kj\"all et~al.}(2014)\citenamefont{Kj\"all, Bardarson,
  and Pollmann}}]{kbp}
\bibinfo{author}{\bibfnamefont{J.~A.} \bibnamefont{Kj\"all}},
  \bibinfo{author}{\bibfnamefont{J.~H.} \bibnamefont{Bardarson}},
  \bibnamefont{and} \bibinfo{author}{\bibfnamefont{F.}~\bibnamefont{Pollmann}},
  \bibinfo{journal}{Phys. Rev. Lett.} \textbf{\bibinfo{volume}{113}},
  \bibinfo{pages}{107204} (\bibinfo{year}{2014}),
  \urlprefix\url{https://link.aps.org/doi/10.1103/PhysRevLett.113.107204}.

\bibitem[{\citenamefont{Luitz et~al.}(2015)\citenamefont{Luitz, Laflorencie,
  and Alet}}]{lla}
\bibinfo{author}{\bibfnamefont{D.~J.} \bibnamefont{Luitz}},
  \bibinfo{author}{\bibfnamefont{N.}~\bibnamefont{Laflorencie}},
  \bibnamefont{and} \bibinfo{author}{\bibfnamefont{F.}~\bibnamefont{Alet}},
  \bibinfo{journal}{Phys. Rev. B} \textbf{\bibinfo{volume}{91}},
  \bibinfo{pages}{081103} (\bibinfo{year}{2015}),
  \urlprefix\url{https://link.aps.org/doi/10.1103/PhysRevB.91.081103}.

\bibitem[{\citenamefont{Peng et~al.}(2018)}]{cappellaro}
\bibinfo{author}{\bibfnamefont{P.}~\bibnamefont{Peng}} \bibnamefont{et~al.},
  \bibinfo{journal}{arXiv preprint arXiv:1901.00034}  (\bibinfo{year}{2018}).

\bibitem[{\citenamefont{Marino et~al.}(2014)\citenamefont{Marino, Majumdar,
  Schehr, and Vivo}}]{majumdar}
\bibinfo{author}{\bibfnamefont{R.}~\bibnamefont{Marino}},
  \bibinfo{author}{\bibfnamefont{S.~N.} \bibnamefont{Majumdar}},
  \bibinfo{author}{\bibfnamefont{G.}~\bibnamefont{Schehr}}, \bibnamefont{and}
  \bibinfo{author}{\bibfnamefont{P.}~\bibnamefont{Vivo}},
  \bibinfo{journal}{Phys. Rev. Lett.} \textbf{\bibinfo{volume}{112}},
  \bibinfo{pages}{254101} (\bibinfo{year}{2014}),
  \urlprefix\url{https://link.aps.org/doi/10.1103/PhysRevLett.112.254101}.

\end{thebibliography}

\end{document}